# Quantitative Determination of Spatial Protein-protein Proximity in Fluorescence Confocal Microscopy

* Yong Wu, *Mansoureh Eghbali, *Jimmy Ou, *Min Li, *[‡§#]Ligia Toro and *[†§#]Enrico Stefani

*Department of Anesthesiology, Division of Molecular Medicine, [‡]Department of Molecular & Medical Pharmacology, [†]Department of Physiology, [§]Brain Research Institute and [#]Cardiovascular Research Laboratory, David Geffen School of Medicine at University of California Los Angeles, Los Angeles, CA 90095-1778

**Address for correspondence**: Yong Wu, David Geffen School of Medicine at UCLA, Department of Anesthesiology, CHS BH-520, BOX 957115, Los Angeles, CA 90095-7115; Tel. 310-794-7804; Fax. 310-825-6649; E-Mail: wuyong@ucla.edu

**To quantify spatial protein-protein proximity (colocalization) in fluorescence microscopic images, cross-correlation and autocorrelation functions were decomposed into fast and slowly decaying components. The fast component results from clusters of proteins specifically labeled and the slow one from image heterogeneity/background. We show that the calculation of the protein-protein proximity index and the correlation coefficient are more reliably determined by extracting the fast-decaying component.**

Colocalization between two fluorescently-labeled proteins, referred here as *protein-protein proximity*, is an essential tool to map and quantify protein-protein interactions. Protein proximity analysis in fluorescence microscopy typically involves a pair of dual color images, in which each color labels one type of protein. A high level of colocalized signals indicates close proximity of the two proteins of interest, which may suggest interactions between them. Among various strategies of colocalization analysis, one of the simplest methods is to overlay the dual color, for example, red and green images, and to assess the amount of overlaid yellow pixels as the indication of interaction [1,2]. Colocalization can also be quantified by various approaches, such as the correlation coefficient (CC)[3,4], the Manders' colocalization coefficients[5-7], the intensity correlation quotient[8], automatic thresholding method[6], and image cross-correlation spectroscopy (ICCS)[7,9,10]. These approaches, however, have their drawbacks. The overlay method is by nature qualitative. Existing quantitative approaches may produce indices without a clear biological meaning, and they often falsely interpret the similarity of spatial pattern in both images (image heterogeneity) as colocalization. Finally, there was no universal way of background reduction. Here we report a novel method that can be applied to a wide range of biological images, successfully minimizing the effect of image heterogeneity and nonspecific fluorescence.

Image intensities are a mixture of specific fluorescence, non-specific fluorescence and random noise, in which specific fluorescence consists of an interacting component and a non-interacting component. The objective of our analysis is to find the fraction of specifically-labeled molecules that generate the interacting component. Our method aims at overcoming two difficulties that most previous methods suffer from. First, the image heterogeneity has to be taken into consideration: Colocalization analysis usually involves calculating covariance, which depends on the spatial structure of cells. In many previous studies, however, either this dependence is neglected, or the analysis has to be restricted within an area in which molecules seem to distribute uniformly. Secondly, the influence of non-specific fluorescence background needs to be minimized. Our method is based on the observation that the cross-correlation value are maximal at $x, y = 0$, decaying as function of $x, y$ pixel shift (cross-correlation function) with sharp and shallow components. We can mathematically show that the sharp component corresponds to the colocalized proteins, while the shallower one to image heterogeneity and the non-specific background. By fitting the sharp and shallow landscape to the sum of two Gaussian functions, one can extract the specific sharp component to calculate the protein proximity index (PPI) values. The Gaussian function was selected to fit the sharp peak because the point spread function (PSF) can be well approximated to this function. The Gaussian function also works fine for the shallow component.

In practice, the method consists of the following steps:

- For a pair of images $I_1$ and $I_2$, calculate the cross-correlation and the autocorrelation $G_{kl}$ ($k,l = 1,2$) as a function of pixel shift $x, y$

$$G_{kl}(x, y) = \frac{\langle I_k(x', y') - \langle I_k \rangle \rangle \langle I_l(x'+x, y'+y) - \langle I_l \rangle \rangle}{\langle I_k \rangle \langle I_l \rangle},$$

where $\langle \rangle$ means calculating the mean value of pixel intensity. Note that in this definition $G_{kl}(x, y)$ can be greater than one.

- For each correlation function, choose a straight line on $x, y$ plane, through which the background drops gently so that the sharp and shallow components can be better distinguished.

- Through the straight line, fit the correlation function values by a sum of two Gaussian functions

$$f(x) = H \cdot e^{-\frac{(x-a)^2}{w^2}} + K \cdot e^{-\frac{(x-a)^2}{s^2}} + C,$$

where $w < s$. When colocalization exists, a successful fit should have $w$ approximately equaling to the full width half maximum of the PSF.

- The estimated PPI values are given by $P_k = \frac{H_{kl}}{H_{ll}}$, ($k = 1,2$ and $k \neq l$), and $CC = \sqrt{P_1 P_2}$, where $H_{kl}$ is the height of fitted peak of the fast component.

This method is very effective in removing the artifacts caused by image heterogeneity. For images with protein clusters, the estimate of PPI values is quite accurate when there is little non-specific component, i.e., when the specific-to-nonspecific ratio (SNR) is high. When the SNR is low, theoretical analysis predicts that the correlation coefficient remains a good estimation, but the PPI values could be greatly distorted by the difference in SNR of the two images. Typical high resolution images show proteins labeled in clusters surrounded by large areas of nonspecific background. In this condition, the median filter background reduction method will estimate the background value at each pixel by calculating the median value of a $n \times n$ square centered at this pixel, with an $n$ at least 10 times larger than the cluster size (in this paper we use a square size of 64×64 or 128×128). This large square size assures that the median value reflects the background level, which can then be subtracted from the image. The resulting image is almost free from nonspecific background.

Compared to existing approaches, our method has the following advantages. First, PPI has the cleaning meaning that they are very close to the fractions of colocalized molecules, if nonspecific labeling is negligible. Secondly, our method is free from false identification of colocalization induced by image heterogeneity. Thirdly, the median-filter method provides a universal and stable approach for background reduction.

The power of our method can be demonstrated by applying it to computer-simulated images, in which the fractions of each type of proteins are exactly known (Supplementary Discussion and Supplementary Fig. 1, Supplementary Table 1 and 2). Our method can produce best estimate to the preset fraction values. For biological images, we initially used a pair of images of a heart cell from mouse where ryanodine receptor (RyR) and estrogen receptor α (ER α) were independently labeled (Fig. 1a and 1b, after median-filter processing). The distribution of proteins in these images clearly forms a spatial pattern along the transverse structures (T-tubules), and the nonspecific fluorescence can be visualized as diffuse labeling surrounding strong specifically labeled protein clusters. Very little colocalization was shown in the overlay (Fig. 1c), contrary to what existing quantitative methods predict (Supplementary Table 3). In Fig. 1d-1i we show correlation functions of the images, and one can see that only autocorrelations show sharp peaks (fast component), while cross-correlation does not, indicating the "colocalization" identified by other methods is not real but merely due to image heterogeneity. This is further confirmed by Fig. 1j-1l, where the nonlinear fit nicely identified the sharp component in autocorrelations, but failed to find it in cross-correlation. Fig. 2 shows the analysis of two images of a mouse heart cell where two different proteins, ryanodine receptor (RyR) and $\alpha_{1C}$ calcium channel ($\alpha_{1C}$), were separately labeled (RyR in Fig. 2a and 2g, and $\alpha_{1C}$ in Fig. 2b and 2h). The overlay of the images (Fig. 2c and 2i) cannot decisively tell whether colocalization exists. The original two images (Fig. 2a and 2b) have very different SNR, resulting unrealistic PPI values: $PPI_{2a}=0.33$ for RyR, $PPI_{2b}=1.21$ for $\alpha_{1C}$. The correlation coefficient CC=0.63 (Fig. 2d-2f). After median filter normalization (Fig. 2g and 2h), PPI values become reasonable: $PPI_{2g}=0.55$ for RyR, $PPI_{2h}=0.76$ for $\alpha_{1C}$ (Fig. 2j-2l). The correlation coefficient is CC=0.64, almost unchanged after median-filter processing, as predicted by theory. Compared to existing methods, the PPI values are lower due to removal of influence from image heterogeneity (Supplementary Table 4). Our method has also been successfully tested on images of a mouse brain cell (astrocyte) where two different proteins, the α subunit of $Ca^{2+}$ and voltage dependent large conductance $K^+$ channels (MaxiK-α) and α-tubulin, were separately labeled (Supplementary Fig. 2), and images in which only a small area of the cell is labeled for MaxiK-α and Thromboxane A2 receptors (TP) on the human kidney cell membrane (Supplementary Fig. 3). In comparison to existing methods (Supplementary Table 5 and 6), the PPI values are considerably lower than other indices, indicting partial colocalization.

To summarize, we have shown that the correlation functions of a pair of fluorescence microscopic images can be decomposed into slow and fast-decaying components, and that the fast-decaying component can be extracted numerically, producing more reliable values of spatial protein proximity index. The described method can serve as a powerful microscopy tool to map and quantify association of macromolecular complexes and their dynamic changes in biological processes.

**ACKNOWLEDGMENTS.**

This work was supported by NIH grants HL088640 (ES), HL054970 (LT) and HL089876 (ME) and AHA Fellowship 0825273F (ML).

**FIGURES**

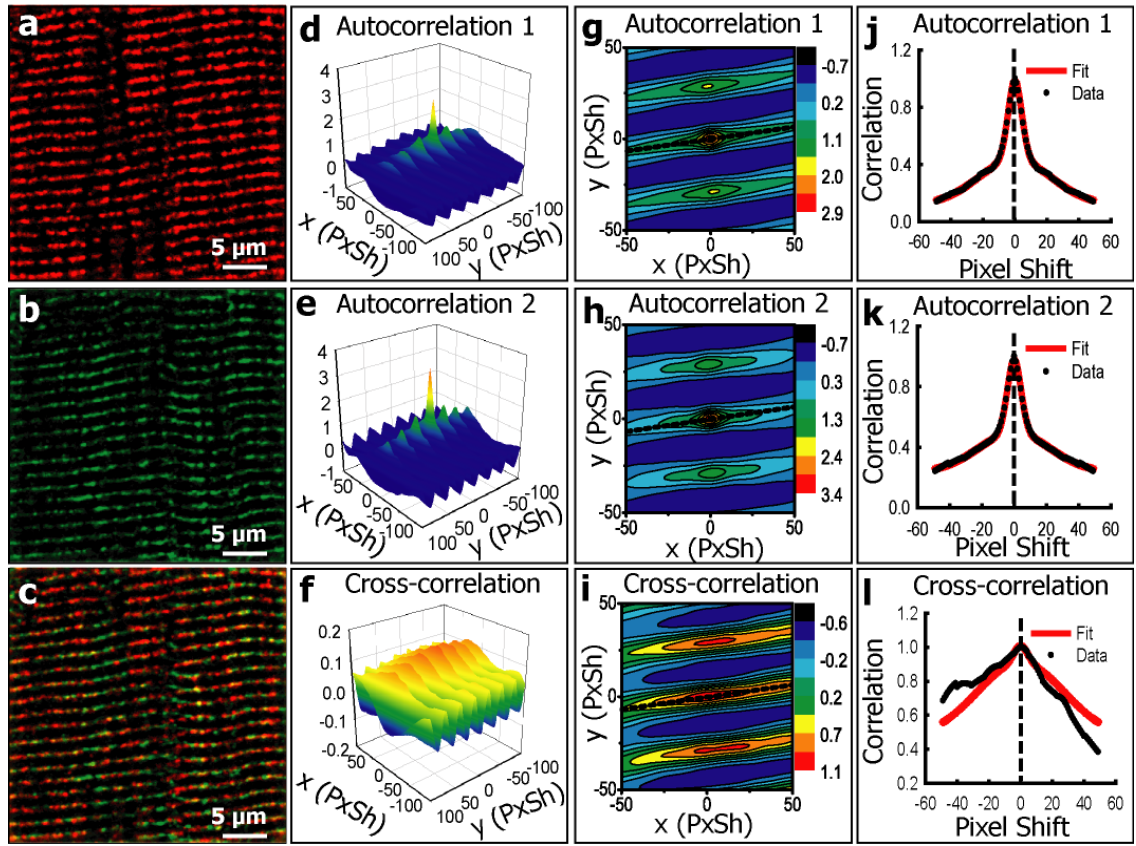

**Figure 1.** Analysis of images of a heart cell from mouse where ryanodine receptor (RyR) and estrogen receptor α (ER α) were independently labeled. (a) RyR channel. (b) ER α channel. (c) Overlay of (a) and (b). (d-f) 3D plot of the cross-correlation and autocorrelation as function of pixel shift (PxSh). (g-i) 2D plot of the cross-correlation and autocorrelation functions, and the line (solid line) through which the nonlinear fit is performed. (j-l) Fitting the cross-correlation and autocorrelation function along the line to the sum of two Gaussian functions. The cross-correlation function does not have a sharp peak, indicting the nonexistence of colocalization.

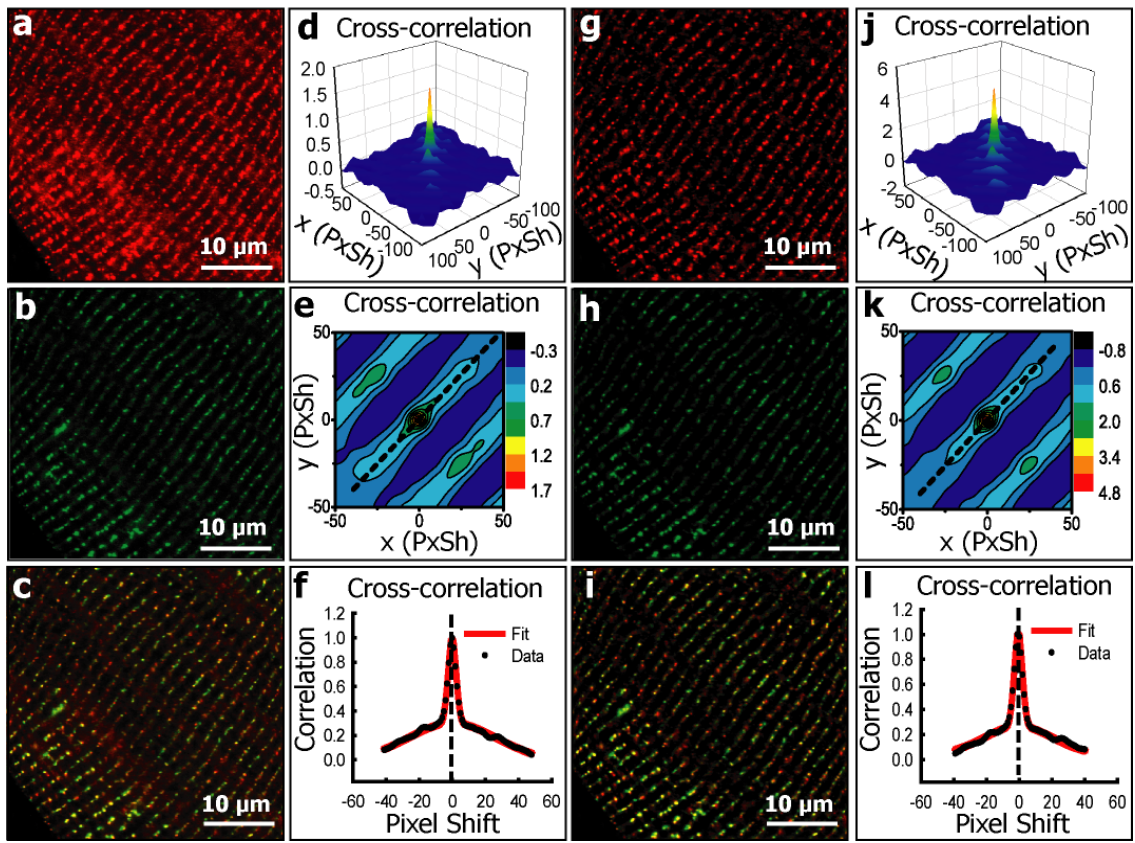

**Figure 2.** Analysis of images of a mouse brain cell (astrocyte) where two different proteins were independently labeled. (a) Ryanodine receptor (RyR). (b) $\alpha_{1C}$ calcium channel ($\alpha_{1C}$). (c) Overlay of (a) and (b). (d-f) The cross-correlation function and the nonlinear fit as described in Fig. 1; PPI is 0.33 for RyR and 1.21 for $\alpha_{1C}$, and CC=0.63. (g-l) Equivalent analysis using median filter background reduction. PPI=0.55 for RyR, PPI=0.76 for $\alpha_{1C}$, and CC=0.64.

# Supplementary Materials

| Supplementary Discussion | Computer Simulated Images and Analysis |
|---|---|
| Supplementary Figure 1 | Colocalization analysis of computer simulated images |
| Supplementary Table 1 | Comparison of colocalization analysis methods for simulated images (Supplementary Fig. 1a and 1b) |
| Supplementary Table 2 | Comparison of colocalization analysis methods for simulated images (Supplementary Fig. 1g and 1h) |
| Supplementary Table 3 | Comparison of colocalization analysis methods on images of a mouse heart cell where ryanodine receptor (RyR) and estrogen receptor α (ER α) were independently labeled (Fig. 1a and 1b) |
| Supplementary Table 4 | Comparison of colocalization analysis methods on images of a mouse heart cell where ryanodine receptor (RyR) and $\alpha_{1C}$ calcium channel ($\alpha_{1C}$) were independently labeled (Fig. 2g and 2h) |
| Supplementary Figure 2 | Analysis of images of a mouse brain cell (astrocyte) where MaxiK-α and α-tubulin were independently labeled |
| *Supplementary Figure 3* | *Colocalization analysis of images of a human embryonic kidney 293T cell (HEK293T) membrane where MaxiK-α and Thromboxane A2 receptor (TPR) were independently labeled* |
| Supplementary Table 5 | Comparison of colocalization analysis methods on images of a mouse brain cell (astrocyte) where MaxiK-α and α-tubulin were independently labeled. (Supplementary Fig. 2g and 2h) |
| Supplementary Table 6 | Comparison of colocalization analysis methods on images of a human embryonic kidney 293T cell (HEK293T) membrane where MaxiK-α and Thromboxane A2 receptor (TPR) were independently labeled (Supplementary Fig. 3g and 3h) |

**Supplementary Discussion: Computer Simulated Images and Analysis**

Computer simulation can generate images with known PPI to test the method. In simulations, the intensity of simulated images is initially set to all zero, and protein clusters are then thrown in as point sources, each generating an intensity distribution according to a Gaussian PSF. The number of proteins can be precisely controlled, and thus the exact PPI values are known. The specifically labeled clusters distinguish themselves from the nonspecifically labeled ones by that they are much brighter. The intensity ratio between a specifically labeled cluster and a nonspecifically label one is set to about 5:1.

Many simulated images have been analyzed and we show two typical examples in Supplementary Fig. 1. We start from a pair of images with a spatial pattern and high specific-to-nonspecific ratio (SNR), shown in Supplementary Fig. 1a-1b, and their overlay in Supplementary Fig. 1c. The real PPI values are set to $P_{1a} \approx 0.20$ and $P_{1b} \approx 0.71$. Supplementary Figure 1d shows the landscape of the cross-correlation function which consists of two clearly distinguishable components, a shallow background reflecting the spatial pattern and a sharp peak on top that accounts for colocalization. The landscape is also shown in Supplementary Fig. 1e as a contour plot, together with a straight line, through which the nonlinear fit is performed. The cross-correlation values through the line can be nicely fitted by the sum of two Gaussian functions, illustrated in Supplementary Fig. 1f. The same procedure can be repeated for the auto-correlation function of each image, and the fitted height of the sharp peak is then used to calculate the estimation of PPI. The result is $P_{1a} \approx 0.22$ and $P_{1b} \approx 0.75$, in excellent agreement with the real values. In Supplementary Table 1, we compare the PPI method to other methods. One can see that previous methods all greatly exaggerate colocalization because of the same spatial pattern the two images have. For example, ICCS gives $P_{1a} \approx 0.56$ and $P_{1b} \approx 0.94$.

To test the method under the influence of nonspecific fluorescence, we add different level of nonspecific background to the two images, shown in Supplementary Fig. 1g and 1h, respectively. The SNR value is 0.16 for Fig. 1g and 7.0 for Fig. 1h, and the real PPI values are unchanged. If background reduction is not performed, our method fails to give reasonable estimate for PPI on this pair, yielding $P_{1g} \approx 0.12$ and $P_{1h} \approx 1.20$, but the correlation coefficient $CC = \sqrt{P_{1g} P_{2h}} \approx 0.38$ is still an excellent estimation (the real value is $\sqrt{0.2 \times 0.71} \approx 0.38$), as predicted by the theory. Median filter can remove most of the background, resulting in visually clearer images shown in Supplementary Fig. 1j and 1k, and their overlay shown in Fig. 1l. The estimated PPI values of median filtered images are $P_{1j} \approx 0.31$ and $P_{1k} \approx 0.49$. These values are close to the real values and capable of signifying the magnitude of colocalization. In Supplementary Table 2, we again compare the PPI method to other methods. One can still see that previous methods usually exaggerate colocalization due to image heterogeneity. We also show that the results vary greatly when different threshold values are used. In this paper, therefore, all images are treated by median-filter background reduction universally rather than by thresholding.

**Supplementary Figure 1: Colocalization analysis of computer simulated images**

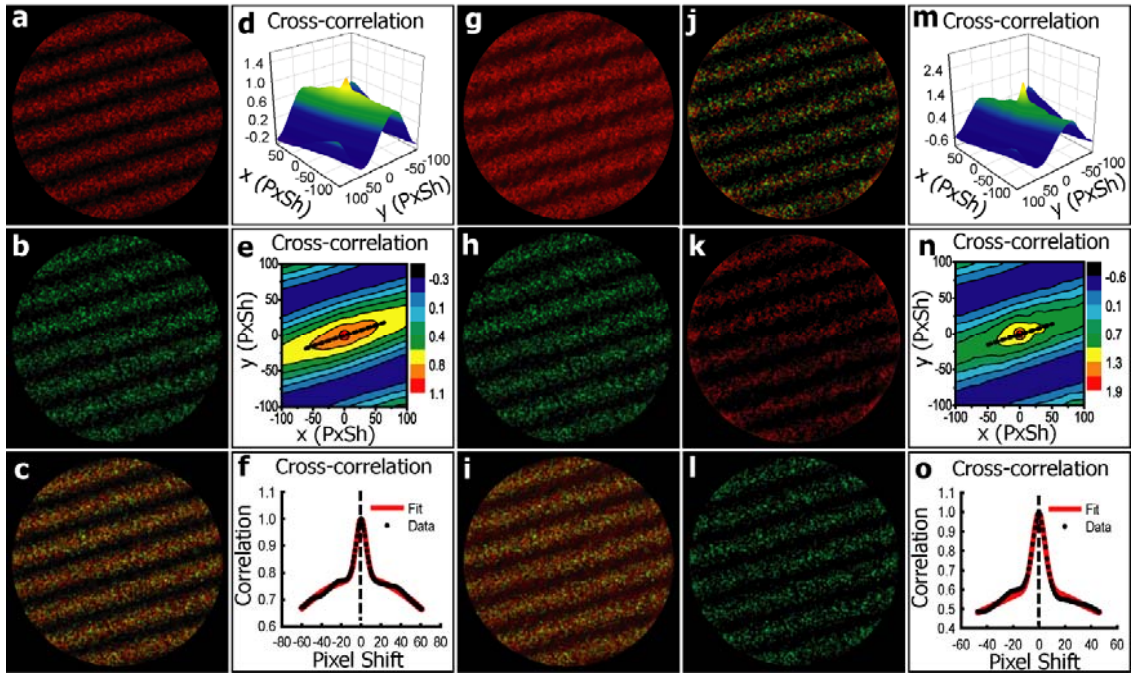

**Supplementary Figure 1** | Colocalization analysis of computer simulated images. **(a-b)** A pair of simulated images with PPI values $P_{1a}=0.20$, $P_{1b}=0.71$, and the correlation coefficient is CC=0.38. In both images the clusters are distributed according to a non-uniform spatial pattern, and the specific-to-nonspecific ratio (SNR) is as high as 10. Naively calculated (without decomposition of the fast and the slow components) PPI values are exaggerated by the spatial pattern: $P_{1a}=0.56$ and $P_{1b}=0.94$. **(c)** Overlay of (a) and (b). **(d)** 3D plot of the cross-correlation function. The peak at the center is due to colocalization and the rest to the non-uniform spatial pattern. **(e)** 2D contour plot of the cross-correlation function. The straight line through the center shows where the double-Gaussian fit is performed. **(f)** Double-Gaussian fit of the cross-correlation function (normalized). The height of sharp peak, together with the heights of the sharp peaks on autocorrelation functions (not shown in this figure), are used to estimate the PPI values. The estimation is excellent: $P_{1a}=0.22$ and $P_{1b}=0.75$. **(g-h)** Simulated images resulted from adding unequal amount of nonspecific background to (a) and (b). The SNR is 0.16 for (g) and 7.0 for (h). **(i)** Overlay of g and h. **(j-k)** Images after the median filter processing. **(l)** Overlay of (j) and (k). **(m)** 3D plot of the cross-correlation function of the median-filtered images. **(n)** 2D contour plot of the cross-correlation function. **(o)** Double-Gaussian fit along the straight line shown in (n). The estimation values are $P_{1j}=0.31$ and $P_{1k}=0.49$, and CC=0.39.

**Supplementary Table 1: Comparison of colocalization analysis methods for simulated images (Supplementary Fig. 1a and 1b)**

|  | Colocalization a to b | Colocalization b to a | Correlation |
|---|---|---|---|
| **Real Value** | 0.20 | 0.71 | 0.38 |
| ***Pearson's Coefficient** | N/A | N/A | 0.73 |
| ***Overlap Coefficient** | 0.13 | 5.11 | 0.83 |
| ***Manders' Coefficient** | 0.98 | 1.00 | N/A |
| **Costes' approach** | 1.00 | 0.96 | N/A |
| **ICCS (image scrambling)** | 0.56 | 0.94 | N/A |
| **PPI** | 0.22 | 0.75 | 0.41 |

\* Calculated by JACoP (http://rsb.info.nih.gov/ij/plugins/track/jacop.html).

**Supplementary Table 2: Comparison of colocalization analysis methods for simulated images (Supplementary Fig. 1g and 1h)**

|  |  | Colocalization g to h | Colocalization h to g | Correlation |
|---|---|---|---|---|
|  | **Real Value** | 0.20 | 0.71 | 0.38 |
| **Median filter** | ***Pearson's Coefficient** | N/A | N/A | 0.54 |
|  | ***Overlap Coefficient** | 0.12 | 3.33 | 0.64 |
|  | ***Manders' Coefficient** | 0.83 | 0.84 | N/A |
|  | **Costes' approach** | 0.72 | 0.68 | N/A |
|  | **ICCS (image scrambling)** | 0.51 | 0.57 | N/A |
|  | **PPI** | 0.28 | 0.47 | 0.39 |
| **Threshold: 1×mean** | ***Pearson's Coefficient** | N/A | N/A | 0.61 |
|  | ***Overlap Coefficient** | 0.084 | 5.42 | 0.70 |
|  | ***Manders' Coefficient** | 0.69 | 0.97 | N/A |
|  | **Costes' approach** | 0.86 | 0.66 | N/A |
|  | **ICCS (image scrambling)** | 0.36 | 0.91 | N/A |
|  | **PPI** | 0.15 | 0.88 | 0.36 |
| **Threshold: 2×mean** | ***Pearson's Coefficient** | N/A | N/A | 0.34 |
|  | ***Overlap Coefficient** | 0.18 | 0.84 | 0.38 |
|  | ***Manders' Coefficient** | 0.72 | 0.36 | N/A |
|  | **Costes' approach** | 0.63 | 0.26 | N/A |
|  | **ICCS (image scrambling)** | 0.54 | 0.21 | N/A |
|  | **PPI** | 0.39 | 0.16 | 0.25 |

\* Calculated by JACoP (http://rsb.info.nih.gov/ij/plugins/track/jacop.html).

**Supplementary Table 3: Comparison of colocalization analysis methods on images of a mouse heart cell where ryanodine receptor (RyR) and estrogen receptor α (ER α) were independently labeled (Fig. 1a and 1b)**

|  |  | Colocalization a to b | Colocalization b to a | Correlation |
|---|---|---|---|---|
| **Median filter** | *Pearson's Coefficient | N/A | N/A | 0.35 |
|  | *Overlap Coefficient | 0.15 | 1.66 | 0.50 |
|  | *Manders' Coefficient | 0.81 | 0.81 | N/A |
|  | Costes' approach | 0.59 | 0.49 | N/A |
|  | ICCS (image scrambling) | 0.32 | 0.37 | N/A |
|  | PPI | 0.08 | 0.06 | 0.07 |
| **Threshold: 2×mean** | *Pearson's Coefficient | N/A | N/A | 0.11 |
|  | *Overlap Coefficient | 0.052 | 0.42 | 0.15 |
|  | *Manders' Coefficient | 0.24 | 0.26 | N/A |
|  | Costes' approach | 0.18 | 0.12 | N/A |
|  | ICCS (image scrambling) | 0.23 | 0.29 | N/A |
|  | PPI | 0.05 | 0.03 | 0.04 |

* Calculated by JACoP (http://rsb.info.nih.gov/ij/plugins/track/jacop.html).

**Supplementary Table 4: Comparison of colocalization analysis methods on images of a mouse heart cell where ryanodine receptor (RyR) and $α_{1C}$ calcium channel ($α_{1C}$) were independently labeled (Fig. 2g and 2h)**

|  |  | Colocalization g to h | Colocalization h to g | Correlation |
|---|---|---|---|---|
| **Median filter** | *Pearson's Coefficient | N/A | N/A | 0.67 |
|  | *Overlap Coefficient | 0.17 | 2.98 | 0.71 |
|  | *Manders' Coefficient | 0.92 | 0.95 | N/A |
|  | Costes' approach | 0.82 | 0.90 | N/A |
|  | ICCS (image scrambling) | 0.54 | 0.83 | N/A |
|  | PPI | 0.55 | 0.76 | 0.64 |

* Calculated by JACoP (http://rsb.info.nih.gov/ij/plugins/track/jacop.html).

**Supplementary Figure 2:** Analysis of images of a mouse brain cell (astrocyte) where MaxiK-α and α-tubulin were independently labeled.

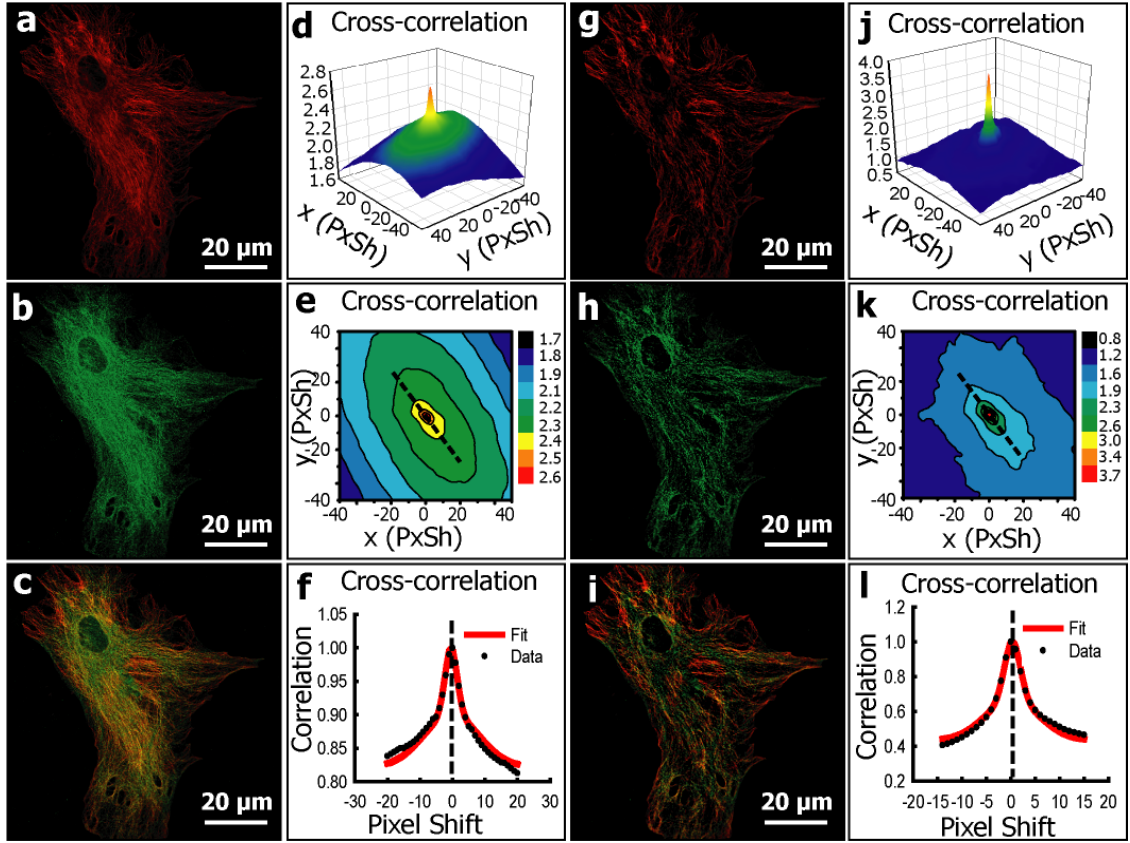

**Supplementary Figure 2:** Analysis of images of a mouse brain cell (astrocyte) where two different proteins were independently labeled. **(a)** MaxiK-α channel. **(b)** α-tubulin. **(c)** Overlay of (a) and (b). **(d-f)** Cross-correlation function and the nonlinear fit as described in supplementary Fig. 1; PPI is 0.56 for MaxiK-α and 0.51 for α-tubulin, and CC=0.54. **(g-l)** Equivalent analysis using median filter background reduction. PPI=0.37 for MaxiK-α, PPI=0.47 for α-tubulin, and CC=0.42.

**Supplementary Figure 3:** Colocalization analysis of images of a human embryonic kidney 293T cell (HEK293T) membrane where MaxiK-α and Thromboxane A2 receptor (TPR) were independently labeled

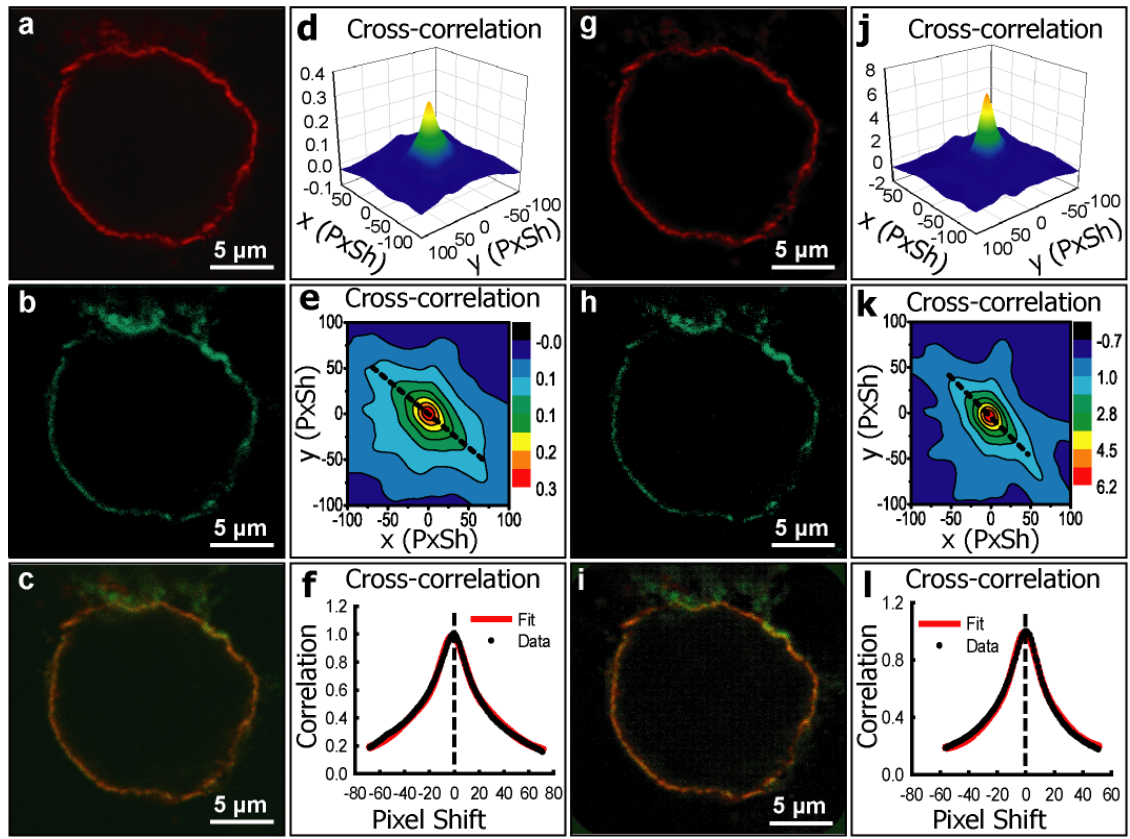

**Supplementary Figure 3:** Colocalization analysis of images of a human embryonic kidney 293T cell (HEK293T) membrane where MaxiK-α and Thromboxane A2 receptor (TPR) were independently labeled. **(a)** MaxiK-α channel. **(b)** TPR channel. **(c)** Overlay of (a) and (b). **(d-f)** Cross-correlation function and the nonlinear fit as described in supplementary Fig. 1; The estimated PPI is 2.29 for MaxiK-α and 0.27 for TPR, and the CC is 0.79. **(g-l)** Equivalent analysis using median filter background reduction. PPI=0.78 for MaxiK-α, PPI=0.64 for TPR, and CC=0.71.

**Supplementary Table 5: Comparison of colocalization analysis methods on images of a mouse brain cell (astrocyte) where MaxiK-α and α-tubulin were independently labeled. (Supplementary Fig. 2g and 2h)**

| | | Colocalization g to h | Colocalization h to g | Correlation |
|---|---|---|---|---|
| Median filter | *Pearson's Coefficient | N/A | N/A | 0.57 |
| | *Overlap Coefficient | 0.64 | 0.62 | 0.63 |
| | *Manders' Coefficient | 0.82 | 0.84 | N/A |
| | Costes' approach | 0.72 | 0.82 | N/A |
| | ICCS (image scrambling) | 0.56 | 0.59 | N/A |
| | PPI | 0.37 | 0.47 | 0.42 |

* Calculated by JACoP (http://rsb.info.nih.gov/ij/plugins/track/jacop.html).

**Supplementary Table 6: Comparison of colocalization analysis methods on images of a human embryonic kidney 293T cell (HEK293T) membrane where MaxiK-α and Thromboxane A2 receptor (TPR) were independently labeled (Supplementary Fig. 3g and 3h)**

| | | Colocalization g to h | Colocalization h to g | Correlation |
|---|---|---|---|---|
| Median filter | *Pearson's Coefficient | N/A | N/A | 0.69 |
| | *Overlap Coefficient | 0.23 | 2.30 | 0.73 |
| | *Manders' Coefficient | 0.92 | 0.94 | N/A |
| | Costes' approach | 0.87 | 0.85 | N/A |
| | ICCS (image scrambling) | 0.79 | 0.69 | N/A |
| | PPI | 0.78 | 0.64 | 0.71 |

* Calculated by JACoP (http://rsb.info.nih.gov/ij/plugins/track/jacop.html).